\documentstyle[prl,aps,twocolumn]{revtex}
\begin{document}
\draft
\title{Gapless Magnetic and Quasiparticle Excitations due to the Coexistence of Antiferromagnetism and Superconductivity in CeRhIn$_5$ : A study of $^{115}$In-NQR under Pressure}
\author{S.~Kawasaki$^1$, T.~Mito$^{1,*}$, Y.~Kawasaki$^{1,**}$, G.~-q.~Zheng$^1$, Y.~Kitaoka$^1$, D. Aoki$^{2,***}$, Y.~Haga$^3$,  and Y.~\=Onuki$^{2,3}$}
\address{$^1$Department of Physical Science, Graduate School of Engineering Science, Osaka University, Toyonaka, Osaka 
560-8531, Japan}
\address{$^2$Department of Physics, Graduate School of Science, Osaka University, Toyonaka, Osaka 560-0043, Japan}
\address{$^3$Advanced Science Research Center, Japan Atomic Energy Research Institute, Tokai, Ibaraki 319-1195, Japan}
\date{\today}
\twocolumn[
\maketitle 
\widetext

\begin{abstract}
We report systematic measurements of ac-susceptibility, nuclear-quadrupole-resonance (NQR) spectrum, and nuclear-spin-lattice-relaxation time ($T_1$) on  the pressure ($P$)- induced heavy-fermion (HF) superconductor CeRhIn$_5$.  The temperature ($T$) dependence of $1/T_1$ at $P$ = 1.6 GPa has revealed that antiferromagnetism (AFM) and superconductivity (SC) coexist microscopically, exhibiting the respective transition at $T_N = 2.8$ K and $T^{MF}_c$ = 0.9 K. It is demonstrated that SC does not yield any trace of gap opening in low-lying excitations below $T_c^{onset} = 2$ K, but  $T_c^{MF} = 0.9$ K, followed by  a $T_1T$ = const law.  These results point to the unconventional characteristics of SC coexisting with AFM.  We highlight that both of the results deserve theoretical work on  the gapless nature in low-lying excitation spectrum due to the coexistence of AFM and SC and the lack of  the mean-field regime below $T_c^{onset} = 2$ K. 
\end{abstract}
\vspace{10mm}

]

\newpage

\narrowtext
In recent years, an intimate interplay between antiferromagnetism (AFM) and superconductivity (SC) has been the most interesting and outstanding issue in cerium (Ce)-based heavy-fermion (HF) systems. The finding of $P$-induced SC in CeCu$_2$Ge$_2$\cite{Jaccard92}, CeIn$_3$, CePd$_2$Si$_2$\cite{Mathur,Grosche} and CeRhIn$_5$\cite{Hegger} strongly suggest that AFM and SC are related to each other since $P$-induced SC occurs either when AFM vanishes or coexists with it \cite{Mito2}. 
Among them, the study on the cubic CeIn$_3$ and the quasi-two dimensional tetragonal CeRhIn$_5$ is promising for systematic investigation of an intimate interplay between AFM and SC. A new HF antiferromagnetic compound CeRhIn$_5$ revealed an antiferromagnetic to superconducting transition at a relatively lower critical pressure $P_c$ = 1.63 GPa and higher $T_c$ = 2.2 K than in previous examples \cite{Jaccard92,Mathur,Grosche,Hegger}.

At $P$ = 0, a previous $^{115}$In-NQR study found that CeRhIn$_5$ has an incommensurate wave vector ${\rm q_M}$ = (1/2, 1/2, 0.297) below the N\'eel temperature $T_N$ = 3.8 K \cite{Curro}.
 A neutron experiment revealed the reduced Ce magnetic moments $M_s$ $\sim$ 0.8$\mu_{\rm B}$ in the antiferromagnetic ordered state \cite{christianson,Bao}.
Our previous NQR study showed that $T_N$ gradually increases up to 4 K as $P$ increases up to  $P$ = 1.0 GPa and decreases with further increasing $P$ \cite{Mito2,Mito,Shinji}. In addition, the temperature ($T$) dependence of nuclear-spin-lattice-relaxation rate $1/T_1$ has probed pseudogap behavior at $P$ = 1.23 and 1.6 GPa \cite{Shinji}. This suggests that CeRhIn$_5$ may resemble other strongly correlated electron systems \cite{Timusk,Kanoda}. 
At $P$ = 2.1 GPa above $P_c$, $1/T_1$ decreases obeying a $T^3$ law without the coherence peak just below $T_c$. This indicates that the SC of CeRhIn$_5$ is unconventional one and it has line-node gap \cite{Mito,Kohori}.
Most remarkably, at $P$ = 1.75 GPa, the onset of AFM is evidenced from a clear split in $^{115}$In-NQR spectrum due to the spontaneous internal field $H_{int}$ below  $T_N$ = 2.5 K. Simultaneously, bulk SC below $T_c$ = 2.0 K is demonstrated by the observation of the Meissner diamagnetism signal. These results have demonstrated that AFM coexists homogeneously with the SC at a microscopic level at the border where both SC and AFM meet one another \cite{Mito2}.

In this letter, we report systematic measurements of ac-susceptibility ($\chi_{ac}$) in $P = 1.12 - 2.0$ GPa and focus on novel superconducting characteristics in the coexistent state of AFM and SC for CeRhIn$_5$ at $P$ = 1.6 GPa.

CeRhIn$_5$ forms in HoCoGa$_5$ type structure where CeIn$_3$ and RhIn$_2$ layers alternately stack along the $c$-axis. Accordingly, there are two inequivalent In sites per unit cell for NQR measurements. The $^{115}$In-NQR measurements at the In(1) site surrounded by four Ce atoms  were performed  using laboratory built spectrometer \cite{Mito2,Curro,Mito,Shinji}.
The NQR spectrum was obtained by plotting the intensity of spin-echo signal as a function of frequency at 1$\nu_{Q}$ ($\pm 1/2\leftrightarrow \pm 3/2$) transition which is the lowest one of the four transitions for the In nuclear spin $I$ = 9/2. The $^{115}$In-NQR $T_1$ was measured at the transitions of 2$\nu_{Q}$ ($\pm 3/2\leftrightarrow \pm 5/2$) above $T$ = 1.4 K, and at 1$\nu_{Q}$ below $T$ = 1.4 K with the saturation-recovery method.
The high-frequency $\chi_{ac}$ measurements under $P$ were carried out by measuring the inductance of an {\it in situ} NQR coil as reported in the previous report\cite{Mito}.  
To obtain hydrostatic $P$, a BeCu piston-cylinder cell was used with Daphne oil (7373) as a $P$-transmitting medium up to $P$ = 2.0 GPa. For our $P$ cell, a value of $P$ distribution ($\Delta P/P$)  is less than 3 \% at low $T$, determined from a broader  NQR linewidth at $P$ = 1.6 GPa than  at ambient $P$.

Fig. 1 indicates a rich $P-T$ phase diagram of CeRhIn$_5$ for AFM and SC referred from the previous report \cite{Mito2}. The SC seems to survive  under AFM near the magnetic criticality in CeRhIn$_5$. Note that the present measurement of $\chi_{ac}$ reveals progressive reduction in the value of bulk superconducting transition temperature $T_c^{MF}$ as shown by closed circle in Fig.1. 

Figs. 2a and 2b indicate the respective $T$ dependencies of $\chi_{ac}$'s and their $T$ derivatives $d\chi_{ac}/dT$ in $P = 1.12 - 2.0$ GPa.  At $P$ = 1.12 GPa,  $\chi_{ac}$ decreases slightly below 1 K with  a size in reduction of $\chi_{ac}$ ($\Delta\chi_{ac}$) due to SC being about only 4\% of $\Delta\chi_{ac}$(bulk) for the bulk Meissner diamagnetism at $P$ = 2.0 GPa. It remains unclear whether this reduction arises from some diamagnetism associated with superconducting fluctuations or possible experimental uncertainty.  At least, it may be ruled out that  bulk superconducting transition does not occur at $P$ = 1.12 GPa.  We note that the specific-heat result under $P$ suggests $P_c(SC)\sim 1.5$ GPa at which the bulk SC sets in \cite{Fisher}.  Consistently,  a bulk nature of  SC in a range $P = 1.53-2.0$ GPa is corroborated by the observation of the Meissner diamagnetism signal whose size is almost the same as in the exclusively superconducting phase. Thus, the SC in CeRhIn$_5$ emerges at pressures exceeding $P_c(SC)\sim 1.5$ GPa as seen in Fig.1.  As seen in Fig. 2a, however, it should be noted that the superconducting transition width  becomes significantly broader with decreasing $P$. A similar behavior is also seen in the $P$ dependence of resistivity over the same $P$ range. Although a zero resistance is observed at $T_c^{zero}$, the decrease in resistance towards $T_c^{zero}$ becomes broadest at $P = 1.6$ GPa (see Fig.5b) \cite{Hegger}. $\Delta\chi_{ac}$'s at $P$ = 1.53 and 1.6 GPa decrease to 80\% of $\Delta\chi_{ac}$(bulk) for the bulk Meissner diamagnetism at $P$ = 2.0 GPa. These anomalous phenomena are associated with  novel superconducting characteristics inherent to the microscopic coexisting state of AFM and SC, but not due to some phase separation between SC and AFM.  This is because the homogeneous AFM over the whole sample is evidenced at $P$ = 1.6 GPa without any trace for the  phase separation between AFM and SC as shown below.  As seen in Fig.2a, $T_c^{onset}$ is defined as a temperature below which the diamagnetism starts to appear, whereas a peak of $d\chi_{ac}/dT$ as $T_c^{MF}$ as seen in Fig .2b. 

Next we present microscopic evidence for novel superconducting characteristics at the coexisting state of AFM and SC in CeRhIn$_5$ at $P$ = 1.6 GPa. The inset of Fig.3 displays the NQR spectra above and below $T_N$ at $P$ = 1.6 GPa. Below $T_N$ = 2.8 K, the NQR spectrum splits into two peaks due to the appearance of $H_{int}$ at the In site. This is clear evidence for the occurrence of AFM at $P$ = 1.6 GPa as well as reported in the previous result at $P$ = 1.75 GPa\cite{Mito2}. The plots of $H_{int}(T)/H_{int}(0)=M_s(T)/M_s(0)$ vs $(T/T_N)$ at $P$ = 0 and 1.6 GPa are compared in Fig.3, showing nearly the same behavior. Here $H_{int}(0)$ is the value extrapolated to $T$ = 0 K and $M_s(T)$ is the $T$ dependence of spontaneous staggered magnetic moment.  
The character of AFM at $P$ = 1.6 GPa is expected to be not so much different from that at $P$ = 0. 

Fig. 4 indicates the $T$ dependence of $1/T_1$ at $P$ = 1.6 GPa. A clear peak in $1/T_1$ is due to critical magnetic fluctuations at $T_N$ = 2.8 K. Below $T_N$ = 2.8 K, $1/T_1$ continues to decrease moderately down to $T_c^{MF}$ = 0.9 K even though passing across $T_c^{onset}\sim 2$ K. This relaxation behavior suggests that the SC does not develop following the mean-field approximation below $T_c^{onset}$.  Most remarkably, $1/T_1$ decreases below $T_c^{MF}$, exhibiting a faint $T^3$ behavior in a narrow $T$ range.  With further decreasing $T$, $1/T_1$ becomes proportional to the temperature, indicative of a gapless nature in low-lying excitation spectrum in the microscopically coexisting state of SC and AFM. Thus the $T_1$ measurement unravels that an intimate interplay between AFM and SC  givies rise to some {\it amplitude fluctuations of superconducting order parameter between $T_c^{onset}$ and $T_c^{MF}$}. 
Such the fluctuations may be responsible for the broad transition in resistance and $\chi_{ac}$ measurements and for the slightly reduced value  less than  $\Delta\chi_{ac}$(bulk) in the $P$ range higher than $P$ = 1.75 GPa (see Fig.2). 
Furthermore, the $T_1T$ = const. behavior well below $T_c^{MF}$  evidences the gapless nature in the coexisting state of AFM and SC. This result is consistent with those in CeCu$_2$Si$_2$ at the border of AFM\cite{Kawasaki1} and a series of CeCu$_2$(Si$_{1-x}$Ge$_2$)$_2$ compounds that show the coexistence of AFM and SC\cite{Kitaoka,Kawasaki2}.  The specific-heat result, that probed a finite value of its $T$-linear contribution, $\gamma_0$ $\sim$100mJ/molK$^2$ at $P$ = 1.65 GPa is now understood due not to a first-order like transition to SC\cite{Fisher}, but to the gapless nature in the coexisting state of AFM and SC.
It is noteworthy that such the $T_1T$ = const. behavior is not observed  below $T_c$ at $P$ = 2.1 GPa\cite{Mito}, consistent with the specific-heat result under $P$ as well \cite{Fisher}. This means that the origin for the $T_1T$ = const. behavior below $T_c^{MF}$ at $P$ = 1.6 GPa is not associated with some impurity effect.  If it were the case, the residual density of states below $T_c$ should not depend on $P$. 
 This novel feature differs from the uranium(U)-based HF antiferromagnetic superconductor UPd$_2$Al$_3$ which has multiple 5$f$ electrons. In UPd$_2$Al$_3$, a superconducting transition occurs at $T_c$ = 1.8 K well below the long-range antiferromagnetic order at $T_N$ = 14.3 K \cite{Geibel,Steglich2}. The $^{27}$Al-NQR $T_1$ results in UPd$_2$Al$_3$ are indicated in the inset of Fig.4 \cite{Tou}.  {\it $1/T_1$ decreases obeying a $T^3$ law over three orders of magnitude below the onset of $T_c$ without any trace for $1/T_1T$ = const. behavior.} This is consistent with the line-node gap even in the coexisting state of AFM and SC below $T_c$. 

In order to highlight the novel superconducting nature in a microscopic level, the $T$ dependence of $1/T_1T$ is shown in Fig.5a at $P$ = 1.6 GPa in $T$ = 0.05 - 6 K and is compared with the $T$ dependence of the resistance $R(T)$ at $P$ = 1.63 GPa referred from the literature \cite{Hegger}. Although each value of $P$ is not exactly the same, they only differ by 2\%.  We remark that the $T$ dependence of $1/T_1T$ points to the pseudogap behavior around $T_{PG}^{NQR}$ = 4.2 K, the AFM at $T_N$ = 2.8 K, and the SC at $T_c^{MF}$ = 0.9 K at which $d\chi_{ac}/dT$ has a peak as seen in Fig.5b. This result itself evidences the microscopic coexistent state of AFM and SC.  A comparison of $1/T_1T$ with  the $R(T)$ at $P=1.63$ GPa in Fig.5b is informative in shedding light on the uniqueness of superconducting and magnetic characteristics. Below $T_{PG}^{NQR}$, $R(T)$ starts to decrease more rapidly than a $T$-linear variation extrapolated from high $T$ side. It continues to decrease across $T_N$ = 2.8 K, reaching zero resistance at $T_c^{zero}\sim$ 1.5 K.
The resistive transition width for SC becomes broader. Unexpectedly, $T_c^{onset}\sim$ 2 K, that is defined as the temperature below which the diamagnetism starts to appear, is higher than $T_c^{zero}\sim$ 1.5 K. Any signature for the onset of SC from the $1/T_1$ measurement is not evident in between $T_c^{onset}$ and $T_c^{MF}$, demonstrating that the mean-field type of gap does not grow up down to $T_c^{MF}\sim$ 0.9 K. {\it The existence of fluctuations due to the interplay of AFM and SC is responsible for the broad transition towards SC that coexists with AFM. }

Finally, we remark that recent neutron-diffraction experiment suggests almost the $P$ independent size of staggered moment $M_s$ in the antiferromagnetic ordered state \cite{Bao}, in contrast to the large $P$ dependence of $H_{int}$ as seen in Fig.1. Its relatively large size of moment with $M_s\sim 0.8\mu_B$ seems to support such a picture that the same $f$-electron exhibits simultaneously itinerant and localized dual nature because there is only one $4f$-electron per Ce$^{3+}$ ion. In this context, it is natural to consider that the superconducting nature in the coexisting state of AFM and SC belongs to a novel class of phase which differs from the {\it conventional} $d$-wave SC with the line-node gap. As a matter of fact, a theoretical model has been recently put forth to address the underlying issue in the coexistent state of AFM and SC\cite{Fuseya}.
 
In conclusion, we have reported the microscopic coexistence of AFM and SC from the systematic measurements of the $T$ dependencies of $1/T_1$ and NQR spectra for CeRhIn$_5$ at $P$ = 1.6 GPa. Also, the coexistent state is suggested to persist down to $P\sim$ 1.5 GPa at least from the measurement of $\chi_{ac}$. 
The detailed measurement of $1/T_1$ has revealed that SC does not yield any trace of gap opening in low-lying excitations below the onset temperature $T_c^{onset}$ = 2 K, but  $T_c^{MF}$ = 0.9 K, followed by the $1/T_1T$ = const. law in low $T$ regime.  These results differ from any previous examples, pointing to the novel characteristics of SC coexisting with AFM.  We highlight that both of the results deserve theoretical work on  the gapless nature in low-lying excitation spectrum due to the coexistence of AFM and SC and the lack of the mean-field regime below $T_c^{onset} = 2$ K. 

One of authors (S.K.) thanks ~K.~Ishida and ~H.~Kotegawa for useful discussions. This work was supported by the COE Research grant (10CE2004) of Grant-in-Aid for Scientific Research from the Ministry of Education, Sport, Science and Culture of Japan.

%\vspace{2cm}
\noindent
* Present address: Department of Physics, Faculty of Science, Kobe University, Nada, Kobe 657-8501, Japan
\noindent
** Present address: Department of Physics, Faculty of Engineering, Tokushima University, Tokushima 770-8506, Japan
\noindent
*** Present address: Institute for Materials Research, Tohoku University, Aobaku Katahira, Sendai 980-8577, Japan

%fig1
\begin{figure}[htbp]
\caption[]{ The $P$-$T$ phase diagram for CeRhIn$_5$.
The respective marks denoted by square, triangle and cross correspond to the pseudogap temperature $T_{PG}^{NQR}$, the antiferromagnetic ordering temperature $T_N$ and the internal field $H_{int}$ at the In site. The open and solid circles correspond to the onset temperature $T_c^{onset}$ and $T_c^{MF}$ of superconducting transition (see text). Dotted line denotes the position for $P=1.6$ GPa. Shaded region indicates coexistent $P$ region of AFM and SC.}
\end{figure}
%fig2
\begin{figure}[htbp]
\caption[]{(a) The $T$ dependencies of $\chi_{ac}$ and (b) the $T$ derivatives of $\chi_{ac}$ ($d\chi_{ac}/dT$) at $P = 1.12-2.0$ GPa.  Arrows indicate $T_c^{MF}$ at each value of $P$ (see text). }
\end{figure}
%fig3
\begin{figure}[htbp]
\caption[]{ Plots of  $H_{int}(T)/H_{int}(0)$ vs $T/T_N$ at $P$ = 0 and 1.6 GPa (see text). Inset shows the $^{115}$In-NQR spectra of 1$\nu_Q$ at $P$ = 1.6 GPa above and below $T_N$ = 2.8 K. }
\end{figure}
%fig4
\begin{figure}[htbp]
\caption[]{The $T$ dependence of $1/T_1$ at $P$ = 1.6 GPa.  Both dotted lines correspond to $1/T_1\propto T$ and $1/T_1\propto T^3$.  Inset indicates the $T$ dependence of $^{27}$Al-NQR $1/T_1$ of UPd$_2$Al$_3$ cited  from the literature \cite{Tou}. Dotted line corresponds to $1/T_1\propto T^3$.}
\end{figure}
%fig5
\begin{figure}[htbp]
\caption[]{(a) The $T$ dependence of $1/T_1T$ at $P$ = 1.6 GPa. (b) The $T$ dependencies of $d\chi_{ac}/dT$ at $P$ = 1.6 GPa and resistance at $P$ = 1.63 GPa cited  from the literature \cite{Hegger}.  $T_c^{MF}$ and $T_c^{onset}$ correspond to the respective temperatures at which $d\chi_{ac}/dT$ has a peak and below which $\chi_{ac}$ starts to decrease. $T_N$ corresponds to the antiferromagnetic ordering temperature at which $1/T_1T$ exhibits a peak and $T_{PG}^{NQR}$ to the pseudogap temperature below which it starts to decrease.
A solid line is an eye guide for the $T$- linear variation in resistance at temperatures higher than $T_{PG}^{NQR}$.}
\end{figure}
\end{document}